\documentstyle[12pt,epsf,fullpage]{article}

\def\be{\begin{equation}}
\def\ee{\end{equation}}
\def\beq{\begin{eqnarray}}
\def\eeq{\end{eqnarray}}

\def\lsim{\:\raisebox{-0.5ex}{$\stackrel{\textstyle<}{\sim}$}\:}
\def\gsim{\:\raisebox{-0.5ex}{$\stackrel{\textstyle>}{\sim}$}\:} 

\begin{document}

\begin{flushright}
TIFR/TH/01-04
\end{flushright}

\vspace*{1in}

\begin{center}
{\Large\bf Looking for the Charged Higgs Boson at
LHC\footnote{Invited talk at the Cairo International Conference on
High Energy Physics, 9-14 January 2001.}} \\[1in]
{\large\bf D.P. Roy} \\[.5cm]
Tata Institute of Fundamental Research \\
Homi Bhabha Road, Mumbai 400 005, India
\end{center}
\bigskip\bigskip

\begin{enumerate}
\item[{}] I discuss LHC signatures of the charged Higgs
boson of the MSSM, focussing mainly on the case of the $H^\pm$ being
heavier than top quark.
\end{enumerate}
\bigskip

The minimal supersymmetric Standard Model (MSSM) contains two complex
Higs doublets, $\phi_1$ and $\phi_2$, corresponding to eight scalar
states.  Three of these are absorbed as Goldstone bosons leaving five
physical states -- the two neutral scalars $(h^0,H^0)$, a
pseudo-scalar $(A^0)$ and a pair of charged Higgs bosons $(H^\pm)$.
All the tree-level masses and couplings of these particles are given
in terms of two parameters, $m_{H^\pm}$ and $\tan\beta$, the latter
representing the ratio of the two vacuum expectation values [1].
While any one of the above neutral Higgs bosons may be hard to
distinguish from that of the Standard Model, the $H^\pm$ carries a
distinctive hall-mark of the SUSY Higgs sector.  Moreover the
couplings of the $H^\pm$ are uniquely related to $\tan\beta$, since
the physical charged Higgs boson corresponds to the combination 
\be
H^\pm = -\phi^\pm_1 \sin\beta + \phi^\pm_2 \cos\beta.
\label{one}
\ee
Therefore the detection of $H^\pm$ and measurement of its mass and
couplings are expected to play a very important role in probing the
SUSY Higgs sector. 

The $t \rightarrow bH^+$ decay is known to provide promising
signatures for charged Higgs boson search at TEVATRON upgrade and LHC
for $M_H < m_t$ [2].  But 
it is hard to extend the $H^\pm$ search beyond $m_t$, because in this
case the combination of dominant production and decay channels, $tH^-
\rightarrow t\bar tb$, suffers from a large QCD background [3].
Moreover the subdominant production channels of $H^\pm W^\mp$ and
$H^\pm H^\mp$ have been found to give no viable signature at LHC [4].
In view of this we have undertaken a systematic study of a heavy
$H^\pm$ signature at LHC from its dominant production channel $tH^-
(\bar tH^+)$, followed by the decays $H^- \rightarrow \bar
tb,\tau\bar\nu$ and $W^- h^0$.  While the 1st represents the dominant
decay channel of charged Higgs boson, the $\tau\nu$ and $Wh^0$ are the
largest subdominant channels in the high and low $\tan\beta$ regions
respectively, with 
\be
B_{\tau\nu} (\tan\beta \gsim 10) \sim 15\% ~~{\rm and}~~ B_{Wh^0}
(\tan\beta = 1 - 5) \lsim 5\%.
\label{two}
\ee
The signature for the dominant decay channel of $H^- \rightarrow \bar
tb$ has been analysed separately assuming three and four $b$-tagging.
The analyses are generally based on parton level Monte Carlo program
with gaussian smearing of lepton and jet momenta for simulating
detector resolution.
\bigskip

\noindent {\bf (i) $H^\pm \rightarrow tb$ Signature with Four $b$-tags
[5]}: 
\medskip

The dominant signal and background processes are
\be
gg \rightarrow tH^- \bar b + {\rm h.c.} \rightarrow t\bar t b\bar b,
\label{three}
\ee
\be
gg \rightarrow t\bar t b\bar b,
\label{four}
\ee
followed by the leptonic decay of one top and hadronic decay of the
other, i.e.
\be
t\bar t b\bar b \rightarrow b\bar b b\bar b W^+ W^- \rightarrow b\bar
b b\bar b \ell\nu q\bar q.
\label{five}
\ee

A basic set of kinematic and isolation cuts,
\be
p_T > 20 ~{\rm GeV}, ~|\eta| < 2.5, ~\Delta R = \left[(\Delta \phi)^2
+ (\Delta \eta)^2\right]^{1/2} > 0.4
\label{six}
\ee
is imposed on all the jet and lepton momenta.  The $p_T$ cut is also
imposed on the missing-$p_T$, obtained by vector addition of the
$p_T$'s after resolution smearing.  This is followed by the mass
reconstruction of the $W$ and the top quark pair, so that one can
identify the pair of $b$-jets accompanying the latter.  While the
harder of these two $b$-jets $(b_1)$ comes from $H^\pm$ decay in the
signal, both of them come mainly from gluon splitting in the
background.  Consequently the $S/B$ ratio is improved by imposing the
following cuts on this $b$-jet pair:
\be
M_{bb} > 120 ~{\rm GeV}, ~E_{b_1} > 120 ~{\rm GeV~~and}~~
\cos\theta_{bb} < 0.75.
\label{seven}
\ee
Then each of this $b$-jet pair is combined with each of the
reconstructed pair of top to give 4 entries for the invariant mass
$M_{tb}$ per event.  One of these 4 entries corresponds to the $H^\pm$
mass for the signal event, while the others constitute a combinatorial
background.  Fig. 1 shows this $tb$ invariant mass distribution for
the signal (\ref{three}) and background (\ref{four}).  The right hand
scale corresponds to the cross-section for $\epsilon^4_b = 0.1$ --
i.e. an optimistic $b$-tagging efficiency of $\epsilon_b = 0.56$.
Reducing it to a more conservative value of $\epsilon_b = 0.4$ would
reduce both the signal and background by a factor of 4 each.

\begin{center}
Table 1. Number of signal and background events \\ in the 4 $b$-tagged
channel per $100 ~{\rm fb}^{-1}$ luminosity \\ in a mass window of
$M_{H^\pm} \pm 40~{\rm GeV}$ at \\ $\tan\beta = 40 ~(\epsilon_b =
0.4)$. 
\end{center}
\[
\begin{tabular}{|c|c|c|c|}
\hline
&&& \\
$M_{H^\pm}({\rm GeV})$ & $S$ & $B$ & $S/\sqrt{B}$ \\
&&& \\
\hline
&&& \\
310 & 32.7 & 26.9 & 6.3 \\
&&& \\
407 & 22.7 & 17.3 & 5.5 \\
&&& \\
506 & 13.2 & ~9.9 & 4.2 \\
&&& \\
605 & ~7.5 & ~5.5 & 3.2 \\
&&& \\
\hline
\end{tabular}
\]
Table 1 lists the number of signal and background events for a typical
annual luminosity of $100~{\rm fb}^{-1}$, expected from the high
luminosity LHC run, assuming $\epsilon_b = 0.4$.  While the $S/B$
ratio is $> 1$, the viability of the signal is limited by the signal
size\footnote{Increasing the $p_T$ cut of $b$-jets from 20 to 30 GeV
would reduce the signal (background) size by a factor of about
\ref{three}(\ref{four} ), hence reducing the viability of this
signal.}.  One expects a $> 3\sigma$ signal upto $M_{H^\pm} = 600~{\rm
GeV}$ at $\tan\beta = 40$. The signal size is very similar at
$\tan\beta = 1.5$, but smaller in between.  The reason the extreme
values of $\tan\beta$ are favoured is that the signal process
(\ref{three}) is controlled by the $tbH^\pm$ Yukawa coupling,
\be
{g \over \sqrt{2} M_W} H^+ \left[\cot \beta m_t \bar t b_L + \tan
\beta m_b \bar t b_R\right],
\label{eight}
\ee
which is large for $\tan\beta \sim 1$ and $\sim m_t/m_b$.
Interestingly these two regions of $\tan\beta$ are favoured by $b -
\tau$ unification for a related reason: i.e. one needs a large
$tbH^\pm$ Yukawa coupling contribution to the RGE to control the rise
of $m_b$ as one goes down from the GUT to the low energy scale [6].
\bigskip

\noindent {\bf (ii) $H^\pm \rightarrow tb$ Signature with Three
$b$-tags [7]:}
\medskip

The contributions to this signal come from (\ref{three}) as well as
\be
gb \rightarrow tH^- + {\rm h.c.} \rightarrow t\bar tb + {\rm h.c.},
\label{nine}
\ee
followed by the leptonic decay of one top and hadronic decay of the
other.  The signal cross-section from (\ref{nine}) is 2-3 times larger
than from (\ref{three}), while their kinematic distributions are very
similar.  Combining the two and subtracting the overlapping piece to
avoid double counting [8] results in a signal cross-section, which is
mid-way between the two.  The background comes from (\ref{four}) as
well as 
\be
gb \rightarrow t\bar tb + {\rm h.c.} ~~{\rm and}~~ gg \rightarrow
t\bar tg,
\label{ten}
\ee
where the gluon jet in the last case is mis-tagged as a $b$-jet.
Assuming the standard mistagging factor of 1\% this contribution turns
out in fact to be the largest source of the background, as we see
below. 

The basic kinematic cuts are as in (\ref{six}) except for a harder
$p_T$-cut, $p_T > 30 ~{\rm GeV}$,
since the 3 $b$-jets coming from $H^\pm$ and $t\bar t$ decays are all
reasonably hard.  This is followed by the mass reconstruction of the
top quark pair as before, so that one can identify the accompanying
(3rd) $b$-jet.  We impose a $p_T > 80 ~{\rm GeV}$ 
cut on this $b$-jet to improve the $S/B$ ratio.  Finally this $b$-jet
is combined with each of the reconstructed top pair to give two
entries of $M_{tb}$ per event.  One of them corresponds to the $H^\pm$
mass for the signal while the other constitutes the combinatorial
background.  Fig. 2 shows this $tb$ invariant mass distribution of the
signal along with the above mentioned backgrounds, including a
$b$-tagging efficiency factor of $\epsilon_b = 0.4$.  
While the $S/B$ ratio is $< 1$ the signal cross-section is much larger
than the previous case.  Table 2 lists the number of signal and
background events for a luminosity of $100~{\rm fb}^{-1}$ at
$\tan\beta = 40$.  The results are very similar at $\tan\beta = 1.5$.
Comparing this with Table 1 we see that the $S/\sqrt{B}$ ratio is very
similar in the two channels.  One should bear in mind however the
larger $p_T$ cut assumed for the 3 $b$-tagged channel.
The cross-sections in both the cases were calculated with the MRS-LO
structure functions [9]. 
\bigskip

\begin{center}
Table 2. Number of signal and background events \\ in the 3 $b$-tagged
channel per $100~{\rm fb}^{-1}$ luminosity \\ in a mass window of
$M_{H^\pm} \pm 40~{\rm GeV}$ at \\ $\tan\beta = 40 ~(\epsilon_b =
0.4)$. 
\end{center}
\[
\begin{tabular}{|c|c|c|c|}
\hline
&&& \\
$M_{H^\pm} ~({\rm GeV})$ & $S$ & $B$ & $S\sqrt{B}$ \\
&&& \\
\hline
&&& \\
310 & 133 & 443 & 6.2 \\
&&& \\
407 & 111 & 403 & 5.6 \\
&&& \\
506 & ~73 & 266 & 4.5 \\
&&& \\
605 & ~43 & 156 & 3.4 \\
&&& \\
\hline
\end{tabular}
\]
\bigskip

\noindent {\bf (iii) $H^\pm \rightarrow \tau\nu$ Signature [10]:}
\medskip

For simplicity we have estimated the signal and background
cross-sections from
\be
gb \rightarrow t H^- + {\rm h.c.} \rightarrow b\bar q q \tau \bar\nu +
{\rm h.c.}
\label{eleven}
\ee
\be
gg \rightarrow t\bar t \rightarrow b\bar q q \bar b\tau\bar\nu + {\rm
h.c.} 
\label{twelve}
\ee
followed by the 1-prong hadronic decay of $\tau$.  The signal contains
a hard a positively polarized $\tau$, while the background contains a
relatively soft and negatively polarized $\tau$ from $W$ boson decay.
The polarization difference can be exploited to sharpen the signal by
simply requiring the charged pion to carry $> 80$\% of the visible
$\tau$ momentum.  Fig. 3 shows the signal and background
cross-sections against the transverse mass of the $\tau$-jet with the
missing-$p_T$.  We see that by exploiting the polarization difference
one can get a clean $H^\pm$ signal in this channel for the large
$\tan\beta$ region at the level of a few fb.  This has been recently
confirmed by a more detailed simulation by the CMS collaboration
including detector acceptance [11].
\bigskip

\noindent {\bf (iv) $H^\pm \rightarrow W^\pm h^0$ Signature [12]:} 
\medskip

We have estimated the signal cross-section from 
\be
gb \rightarrow tH^- + {\rm h.c.} \rightarrow bW^+ W^- h^0 + {\rm
h.c.},
\label{fourteen}
\ee
followed by $h^0 \rightarrow b\bar b$, $W^\pm \rightarrow \ell\nu$ and
$W^\mp \rightarrow q\bar q$.  Thus the final state consists of the
same particles as the dominant decay mode of eq. (\ref{nine}).
Therefore we have to consider the background from the $H^- \rightarrow
t\bar b$ decay (\ref{nine}) along with those from the QCD processes of
eq. (\ref{ten}). 

We require 3 $b$-tags along with the same basic cuts as in section
(ii).  This is followed by the mass reconstruction of $W^\pm$ and the
top, which helps to identify the accompanying $b$-pair and the $W$.
The resulting $bb$ and $Wb$ invariant masses are then subjected to the
constraints,
\be
M_{bb} = m_{h^0} \pm 10~{\rm GeV} ~{\rm and}~ M_{Wb} \neq m_t \pm 20
~{\rm GeV}.
\label{fifteen}
\ee
The $h^0$ mass constraint and the veto on the second top helps to
distinguish the $H^\pm \rightarrow W^\pm h^0$ signal from the
backgrounds.  However the former is severely constrained by the signal
size as well as the $S/B$ ratio.  Consequently one expects at best a
marginal signal in this channel and that too only in a narrow strip of
the $M_{H^\pm} - \tan\beta$ parameter space, at the boundary of the
LEP exclusion region.  Fig. 4 shows the signal (\ref{fourteen}) along
with the backgrounds from (\ref{nine}) and (\ref{ten}) against the
reconstructed $H^\pm$ mass at one such point -- $M_{H^\pm} = 220~{\rm
GeV}$ and $\tan\beta = 2$.

The LEP limit of
$m_{h_0} (m_{A_0}) > 100 ~{\rm GeV}$ 
in the low $\tan\beta$ region implies that the $H^\pm \rightarrow Wh^0
(WA^0)$ decay channel has as high a threshold as the $t\bar b$ channel,
while the latter has a more favourable coupling.  Consequently the
$H^\pm \rightarrow Wh^0 (WA^0)$ decay $BR$ is restricted to be $\lsim
5\%$ over the LEP allowed region [13].  However the LEP constraint
does not hold in singlet extensions of the MSSM like
the NMSSM [14].  Consequently the $H^\pm \rightarrow Wh^0 (WA^0)$ can
be the dominant decay mode for $M_{H^\pm} \sim 160~{\rm GeV}$ in the
low $\tan\beta$ region and lead to a spectacular signal at the LHC, 
as is illustrated in [12].  One needs a systematic analysis of the
LHC signatures for both neutral and charged Higgs bosons in the NMSSM,
particularly in the low $\tan\beta$ region.

It should be mentioned here that these parton level Monte Carlo
analyses of the $H^\pm$ signature in $tb$ and $Wh^0$ decay channels
need to be followed up by detailed simulation with PYTHIA, including
detector acceptance, as in the case of the $\tau\nu$ channel [11].  
Finally it should be
noted that practically all the analyses of $H^\pm$ signal so far are
based on the lowest order production vertex, represented by the Yukawa
coupling of eq. (\ref{eight}).  One loop electroweak corrections to
this vertex can give upto 10 (20)\% reduction in the signal
cross-section at high (low) $\tan\beta$, as recently shown in [15].
The corresponding correction from QCD loops is expected to be larger,
but not yet available.  This is evidently important for a quantitative
evaluation of this signal.  One should also bear in mind the
possibility of a large correction to this vertex from SUSY loops
depending on the choice of SUSY parameters [16].

\newpage

\hrule width 0pt
\vspace*{1in}

\includegraphics{looking1.eps} 
\label{fig:looking1}
\vspace*{1.3in}
\begin{enumerate}
\item[{}] Fig. 1. The reconstructed $tb$ invariant mass distribution of
the $H^\pm$ signal (\ref{three}) and the QCD background (\ref{four})
in the isolated lepton plus multi-jet channel with 4 $b$-tags.  The
scale on the right corresponds to a $b$-tagging efficiency fator
$\epsilon^4_b = 0.1$.
\end{enumerate}


\vspace*{1in}

\includegraphics{looking2.eps} 
\label{fig:looking2}
\vspace*{1.6in}
\begin{enumerate}
\item[{}] Fig. 2. 
The reconstructed $tb$ invariant mass distribution of
the $H^\pm$ signal and different QCD backgrounds in the isolated
lepton plus multijet channel with 3 $b$-tags.
\end{enumerate}

\newpage

\hrule width 0pt

\includegraphics{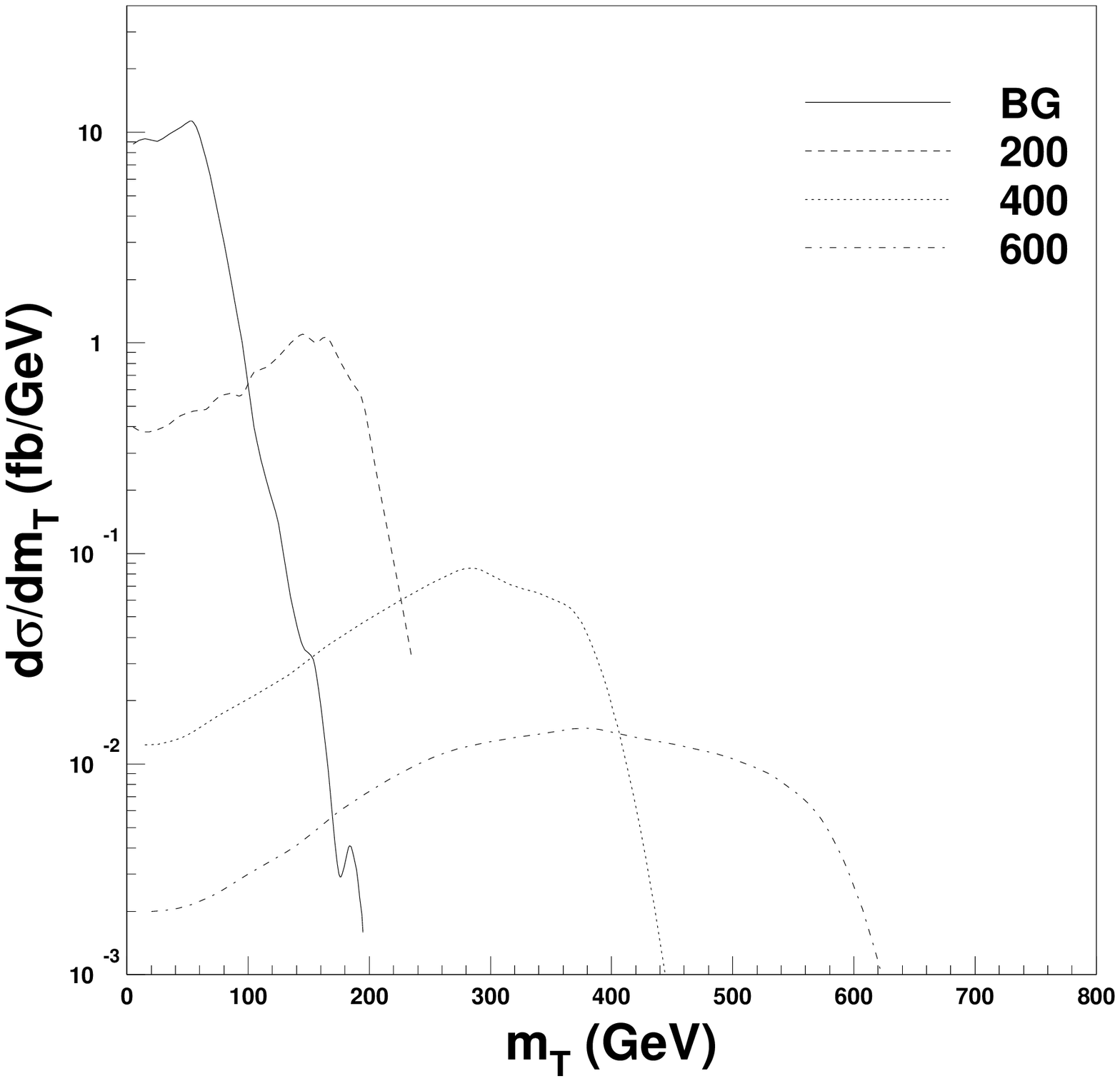} 
\label{fig:looking3a}
\includegraphics{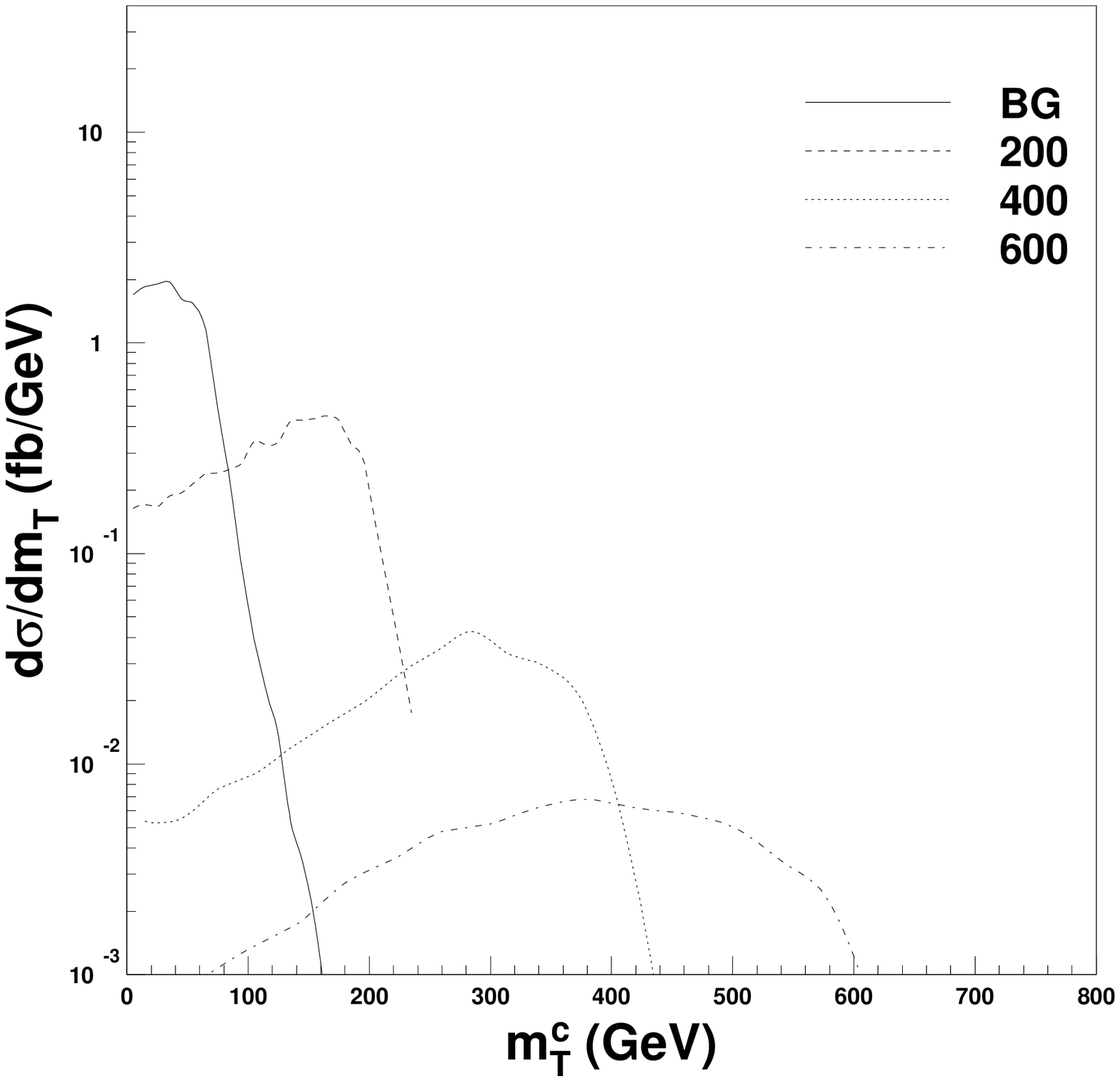} 
\label{fig:looking3b}
\vspace {3in}
\begin{enumerate}
\item[{}] Fig. 3. Distribution of the $H^\pm$ signal and background
cross-sections in the transverse mass of the $\tau$-jet with the
missing-$p_T$ for (a) all 1-prong $\tau$-jets, (b) those where the
charged pion carries over 80\% of the $\tau$-jet $p_T$ $(M_{H^\pm} =
200,400,600 \ {\rm GeV \ and} \ \tan\beta = 40)$.
\end{enumerate}

\includegraphics{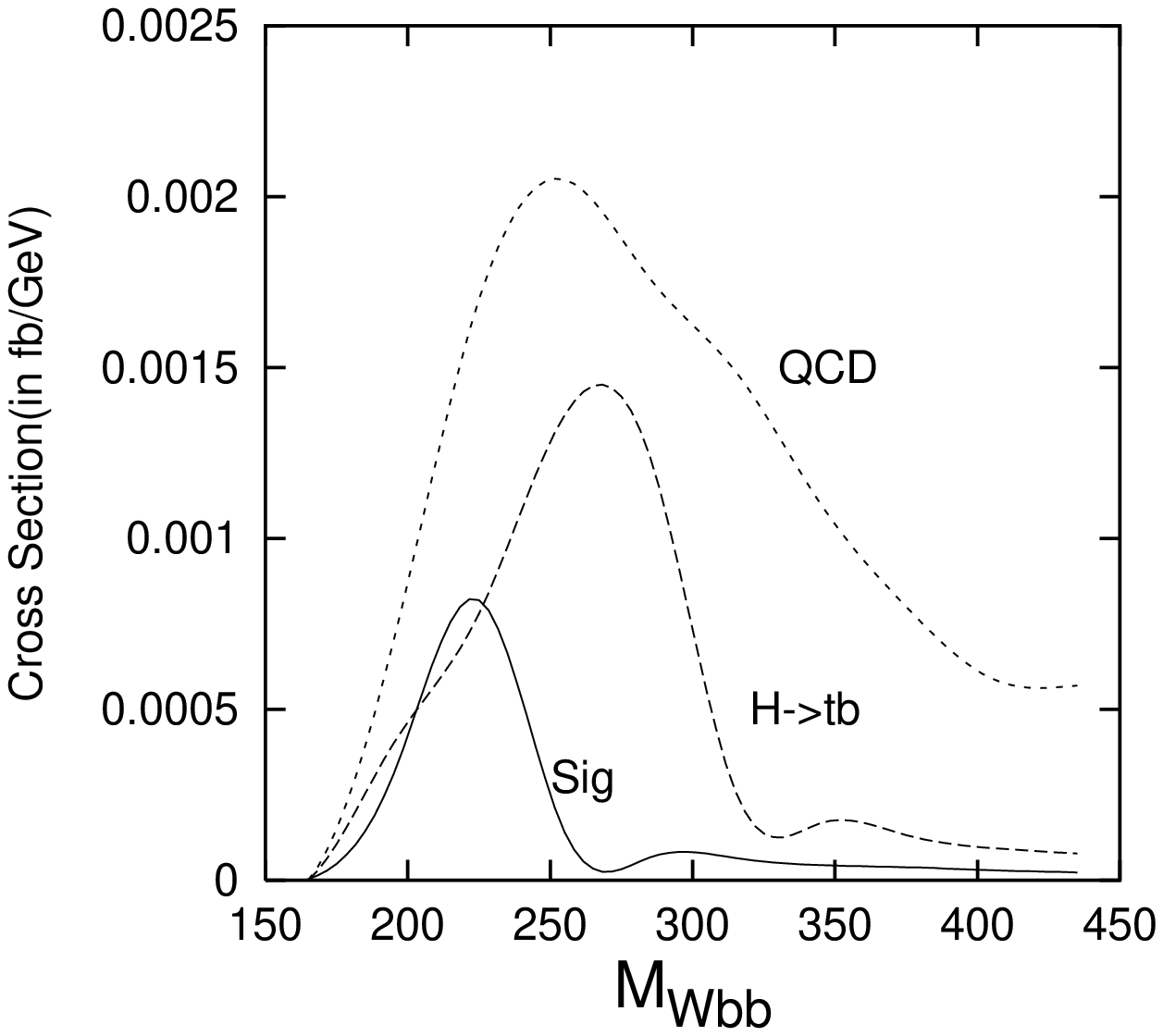} 
\label{fig:looking4}
\vspace{1.6in}
\begin{enumerate}
\item[{}] Fig. 4. The $H^\pm \rightarrow Wh^0$ signal cross-section is
shown against the reconstructed $H^\pm$ mass for $M_{H^\pm} = 220~{\rm
GeV}$ and $\tan\beta = 2$ along with the $H^\pm \rightarrow tb$ and
the QCD backgrounds.
\end{enumerate}
\end{document}